\documentclass{article}

\usepackage{amsmath,amssymb,amsthm}
\usepackage{psfrag}
\usepackage{bm}
\usepackage{graphics}
\usepackage{epsfig}

\newcommand{\vc}{\mathbf}
\newcommand{\bs}{\boldsymbol}
\newcommand{\bb}{\boldsymbol{\beta}}

\newtheorem{proposition}{Proposition}[section]
\newtheorem{lemma}{Lemma}[section]

\begin{document}


\title{A Fast Algorithm for Robust Regression with  Penalised Trimmed Squares}

\author{L. Pitsoulis and G. Zioutas \\
       Department of Mathematical and Physical Sciences \\ 
                     Aristotle University of Thessaloniki \\
                     541 24 Thessaloniki, Greece }

\maketitle

\begin{abstract}
\noindent
The presence of groups containing high leverage outliers makes linear regression  a difficult problem due  to the masking effect. 
The available high breakdown estimators based on Least Trimmed Squares  often do not succeed in detecting masked high 
leverage outliers in finite samples. 

An alternative to the LTS estimator, called Penalised Trimmed Squares (PTS) estimator, was introduced by the authors 
in~\cite{ZiouAv:05,ZiAvPi:07} and it appears to be less sensitive to the masking problem. This estimator 
is defined by a Quadratic Mixed Integer Programming (QMIP) problem, where in the objective function a penalty cost for
each observation is included which serves as an upper bound on the residual error for any feasible regression line. 
Since the PTS does not require presetting the number of outliers to delete from the data set, it has better efficiency
with respect to other estimators. 
However, due to the high computational complexity of the resulting QMIP problem, exact solutions for moderately 
large regression problems is infeasible. 

In this paper we further establish the theoretical properties of the PTS estimator, such as high breakdown and efficiency, 
and propose an approximate algorithm called Fast-PTS to compute the PTS estimator for large data sets efficiently. 
Extensive computational experiments on sets of benchmark  instances with varying degrees of outlier contamination, 
indicate that the proposed algorithm performs well in identifying groups of high leverage outliers in reasonable computational
time.

\vspace*{.25in}
\end{abstract}

\setcounter{tocdepth}{4}
\tableofcontents

\section{Introduction}  \label{Sec1}              

In multilinear regression models experimental data often contains outliers and bad influential observations, 
due to errors. It is important to identify these observations and eliminate them from the data set, since
they can lead the regression estimate to take erroneous values.  If the data is contaminated with a single or few outliers the problem of 
identifying such observations is not difficult, but in most cases data sets contain groups of outliers which makes the problem more
difficult due to masking and swamping effects. 
An indirect approach to outlier identification is through a robust regression estimate. If a robust estimate is relatively 
unaffected from outliers, then the residuals from the robust fit should be used to identify the outliers.

Two commonly used criteria for robustness are the {\em breakdown point} and {\em efficiency} of the estimator. 
The breakdown point can be roughly defined as the minimum percentage of outliers present in the data, that could 
lead the error between the  robust estimate  
and the hypothetical true estimate to be infinitely large. It is desirable for a robust estimator to have a breakdown
point of close to 50\% of the sample size.

A well known  estimator with high breakdown point is the Least Trimmed Squares (LTS) estimator of Rousseeuw and 
Leroy~\cite{RouLer:87}. The method consists of finding a ${n-k}$ subset of observations whose deletion from the data set would 
lead to the $k$ smallest residual sum of squares. However, it is well known that the LTS loses efficiency.
Several robust estimators have been proposed in the literature as extensions to the LTS, 
to obtain high breakdown point and simultaneously improve the efficiency. 
Among them are the S-estimator proposed by Rousseeuw and Yohai~\cite{RouYoh:84}, the MM-estimator proposed by 
Yohai~\cite{Yohai:87}, the $r-$estimator proposed by Yohai and Zamar~\cite{YohZam:88}, the S1S estimator proposed by  
Coakley and Hettmansperger~\cite{CoaHet:93}, WLSE computed from an initial high breakdown robust estimator by 
Agostinelli and Markatou~\cite{AgoMar:98} and the REWLS by Gervini and Yohai~\cite{GerYoh:02}. Most of these estimators 
start from an initial robust scale $\sigma_{LTS}$ and robust regression coefficient estimate $\bb_{LTS}$ given 
from the LTS estimator with coverage $k\simeq[(n+p+1)/2]\simeq 50\%$. Then they compute the standardised residuals 
using different schemes based on robust scale $\sigma_{LTS}$ and design weights  
with appropriate cut-off values, and they reconsider which of the potential outliers should remain in the basic subset as clean 
data and which should be eliminated as outliers. Thus, they simultaneously attain the maximum breakdown point of the LTS
estimator while improving its efficiency. 
The key to the success of all the aforementioned  methods is to start with a good initial 
regression coefficient estimate $\bb_{LTS}$. When a finite sample contains multiple high leverage points, 
these may be masked outliers and would affect the LTS estimator resulting in a biased initial estimate.
Moreover, the LTS method requires a priory knowledge of the {\em coverage} $k$, or equivalently the number $n-k$ of the most likely outliers that 
produces the largest reduction in the residual sum of squares when deleted. Unfortunately, this knowledge of $k$ is 
typically unknown, Gentleman and Wilk~\cite{GenWilk:75}. Computation of the LTS estimator requires the solution of 
a hard combinatorial problem, and there have been many exact and approximation algorithms proposed in the literature
(see Section~\ref{subsec_lts}).

A different approach to obtain a robust estimate has been proposed in~\cite{ZiAvPi:07} called Penalised Trimmed Squares
(PTS), which does not require presetting the number $n-k$ of outliers to delete from the data set. The PTS approach ``trims'' 
outliers from the data but instead of discarding a fixed number of observations, a fixed threshold for the allowable size 
of the adjusted residuals is used. The new estimator PTS is defined by minimising a {\em loss function}, 
which is the sum of squared residuals and penalty costs for deleting  bad observations. These 
penalties regulate the threshold of the allowable adjusted residuals, as well as the coverage. In order to overcome the
problem of groups of masking outliers containing almost identical high leverage points, lower penalties  are proposed
yielding a smaller adjusted residual threshold for such observations. These penalties are a function of robust 
leverages resulted from the MCD estimator by Rousseeuw and Van Driessen~\cite{RouDrie:99}. 
Computationally,  the PTS estimator also involves the solution of a hard combinatorial problem.

The purpose of this paper is twofold. First, the robust properties of the PTS estimator as presented in~\cite{ZiAvPi:07}
such as equivariance, exact fit property, high breakdown point and efficiency, 
are established. Secondly we present an efficient approximation algorithm called Fast-PTS,  
to compute the PTS estimator for large data sets without having to solve the problem exactly. 

The organisation of the paper is the following. Section~\ref{Sec2} contains some preliminary notations and definitions. 
The definition of the PTS estimator, its robust properties, the computation of the penalties and  finally an equivalent
mixed integer quadratic programming formulation are presented in  Section~\ref{sec_pts}. Section~\ref{sec_fast-pts} describes
the Fast-PTS algorithm, where initially a set of necessary optimality conditions is given, and the algorithm is then presented
in pseudocode. Extensive computational experiments to examine the performance of the Fast-PTS algorithm with respect to,
both the quality of the approximate solution when compared to the exact solution of small instances, and the robust 
performance to a set of benchmark and artificially generated large instances, are given in Section~\ref{sec_comp}.
Finally, concluding remarks and further possible extensions to both the algorithm and the robust estimator are given
in the last section.

\section{Trimmed Squares Regression}\label{Sec2}

\subsection{Preliminaries}
In this section we will state some well established facts in order to set up the notation that will be used for the rest of 
the paper. 
We consider the multi-linear regression model with $p$ independent variables
\begin{equation} \label{eq_1}
\vc{y} = \vc{X}{\bs \beta} + \vc{u},
\end{equation}
where $\vc{y} = (y_{1},y_{2}, \ldots , y_{n})^{T}$ is the $n\times 1$ vector of the response variable, 
$\vc{X}$ is a full rank $n\times p$ matrix with rows 
$\vc{x}_{i} =(x_{i1},x_{i2},\ldots ,x_{ip})$ of explanatory variables, 
$\bs{\beta} $ is a $p\times 1$ 
vector  $\bs{\beta} =(\beta _{1}, \beta _{2}, \ldots , \beta _{p})^{T}$  of  unknown parameters , 
and $\vc{u}$ is a $n\times 1$ 
vector $\vc{u} = (u_{1}, u_{2}, \ldots , u_{n})^{T}$ of iid random errors with expectation zero and variance $\sigma^2$. 
We observe a sample $(x_{i1}, x_{i2}, \ldots ,x_{ip}, y_{i})$, for $i=1, 2, \ldots , n$, and construct an estimator 
for the unknown parameter $\bb$. The Ordinary Least Squares Estimator (OLS) is defined by minimising the squared residual 
loss function 
\begin{equation}\label{eq_2}
OLS(\vc{X,y}) := \arg\min_{\bb} \sum_{i=1}^{n}r(\vc{\bb})^{2}_{i}
\end{equation}
where $r(\bb)_{i}$ is the regression residual $r(\bb)_{i}:=y_{i}-\vc{x}_{i}{\bb}$. We will write 
$r_{i}$ instead of $r(\bb)_{i}$ whenever the parameter vector $\bb$ need not be explicitely
stated. It is well known that a solution to (\ref{eq_2}) is obtained in polynomial time (i.e. ${\cal O}(n^3)$) by
the normal equations
\begin{equation*}
OLS(\vc{X,y}) = (\vc{X}^{T}\vc{X})^{-1}\vc{X}^{T}\vc{y}
\end{equation*}

A transformation of the residuals that will be useful in this work is the adjusted residual $\alpha_{i}$ which is defined as 
\begin{equation}\label{eq_adj_residual}
\alpha_{i}:=\frac{r_{i}}{\sqrt{1-h_{i}}}, 
\end{equation}
where $h_{i} \;\;(0<h_{i}<1)$ measures the leverage of the $i^{th}$ observation and is defined as
$h_{i} := {\vc{x}_{i}^{T}}(\vc{X}^{T}\vc{X})^{-1}\vc{x}_{i}$.
The reduction in the sum of squared residuals in (\ref{eq_2}) due to the deletion of an observation 
$(\vc{x}_{i},y_{i})$ is the square of its adjusted residual
\[
\alpha_{i}^2=\frac{r_{i}^{2}}{1-h_{i}}.
\]
Unfortunately, points that are far from the predicted line (outliers) are overemphasised. There are several types of 
outliers that can occur in a data set $(\vc{X,y})$. 
Following the terminology of Rousseeuw and Van Zomeren~\cite{RouZom:90}, a point $(\vc{x}_{i}, y_{i})$ which 
does not follow the linear pattern of the majority of the data but whose $\vc{x}_{i}$ is not outlying is called a 
{\em vertical outlier}. A point $(\vc{x}_{i}, y_{i})$ whose $\vc{x}_{i}$ is outlying (large $h_{i}$) is called a 
{\em leverage point}. It is called {\em good} leverage point when $(\vc{x}_{i}, y_{i})$ follows the pattern of the 
majority otherwise it is called {\em bad} leverage point. The OLS estimator is very sensitive to outliers,
in the sense that the presence of any type of previously mentioned types of outliers greatly affects the solution
of (\ref{eq_2}). 
We wish to construct a robust estimator for the parameter $\bs{\beta}$, such that the influence of any 
observation $(\vc{x}_{i}, y_{i})$ on the sample estimator is bounded.

\subsection{Least Trimmed Squares}\label{subsec_lts}

Rousseeuw introduced the Least Trimmed Squares (LTS) estimator in~\cite{Rousseeuw:84}, which fits the best subset of 
$k$ observations, removing the rest $n-k$ observations. The LTS estimator is defined as follows
defined as:
\begin{eqnarray}\label{eq_LTS}
LTS(\vc{X,y},k):=  \arg\min_{\bb} & &  \sum_{i=1}^{k}  r(\bb)_{(i)}^2 \\ \nonumber
                \mbox{s.t.}     & &   r(\bb)_{(1)}^{2} \le r(\bb)_{(2)}^{2} \le \cdots \le r(\bb)_{(n)}^{2} 
\end{eqnarray}
where $k$ is called the {\em coverage}, and is chosen as $k\geq [(n+p+1)/2]$ a priori, so as to maximise the
breakdown point. In (\ref{eq_LTS}) and in what follows, we employ the convention of writing $r(\bb)_{(i)}$
for the $i^{\mbox{th}}$ smallest residual error with respect to $\bb$. 
The LTS estimator has high breakdown point but loses efficiency, since $n-k$ observations have to be removed from the 
sample even if they are not outliers. Assuming that the set of points $(\vc{x}_{i},y_{i})$ are in {\em general 
position}, that is any two distinct subsets of $p$ points have different regression lines, and observing that for 
a given $k$ the problem in (\ref{eq_LTS}) consists of solving ${n \choose k}$ ordinary least squares problems, it is 
immediate that it has a unique solution. 

Problem (\ref{eq_LTS}) has been formulated as a nonlinear programming problem with linear constraints and
non-convex objective function by Giloni and Padberg in~\cite{GilPad_a:02}, thereby its local minima are
well characterised by the Karush-Kuhn-Tucker optimality conditions. In particular the authors in~\cite{GilPad_a:02}
establish its equivalency with a concave minimisation problem, which is known to be NP-Hard, 
and provide a procedure for computing local minima. 
The exact computation of (\ref{eq_LTS}) requires an exponential number of steps with respect to the size 
of the problem, therefore it can only be applied to small instances. Agull\'{o} in~\cite{Agullo:01} presented 
a branch and bound procedure for the exact computation of (\ref{eq_LTS}) along with several procedures to 
reduce the computational effort, and managed to find the global optimum for several benchmark instances 
of size $n<50$. 
Polynomial time algorithms for the exact computation of the LTS for simple ($p=1$) linear regression problems are given
by H\"ossjer~\cite{Hossjer:95}, and recently by Li~\cite{Li:05} who utilises advanced data structures to attain
improved computational complexity.
Due to the high computational complexity of the LTS estimator there
have been a plethora of  approximation algorithms appeared in the literature with varying success. 
We mention among others, the Forward Search algorithm by Atkinson~\cite{Atkinson:94,AtkChen:94}, 
the Feasible Solution algorithm by Hawkins~\cite{Hawkins:94,HawOli:99} and the more recent Fast-LTS algorithm
by Rousseeuw and Van Driessen~\cite{RouDrie:06} which currently the most accurate and efficient algorithm.


\section{Penalised Trimmed Squares}\label{sec_pts}

The basic idea of the PTS estimator is to insert fixed penalty costs for each observation into the objective function of 
(\ref{eq_LTS}), 
such that only observations whose adjusted residuals is larger than their penalty costs are deleted from the data set. 
Therefore instead of defining a priori a coverage $k$, the penalty costs are defined and the coverage becomes a decision
variable. In the sections that follow suitable penalties for multiple high-leverage outliers are proposed.
We consider as most likely outliers that subset of the observations that produces significant reduction in the sum of squared residuals 
when deleted. The proposed PTS estimator can be defined through a minimisation problem, with an objective function
split into two parts; the sum of $k$ squared residuals in the clean data and the sum of the penalties for deleting the rest 
$n-k$ observations. 
\begin{eqnarray}
PTS(\vc{X,y},\vc{p}):= \arg\min_{\bb,k}& & 
\sum_{i=1}^{k} r(\bb)_{(i)}^{2} + \sum_{i=k+1}^{n} p_{(i)}  \label{eq_PTS} \\\nonumber
\mbox{s.t.}             & & r(\bb)_{(1)}^{2} \le r(\bb)_{(2)}^{2} \le \cdots \le r(\bb)_{(n)}^{2} 
\end{eqnarray}
where $\vc{p} = (p_{1},\ldots,p_{n})$  and $p_{i}$ is the {\em penalty} cost for deleting the $i^{\mbox{th}}$ observation defined as
\[
p_{i} := \max \{ \epsilon, (c\hat{\sigma})^{2} \}
\]
for some small $\epsilon > 0$, 
where $\hat{\sigma}$ is a {\em robust residual scale} and $c$ is the {\em cut-off parameter}.
Although these penalties are constant in the basic definition of the PTS, as we shall see later in section~\ref{subsec_penalties}
each observation will have an individual penalty. 
The estimator performance is very sensitive to the penalties defined a priori which regulate 
the robustness and the efficiency of the estimator. The choice of the robust scale $\hat{\sigma}$ plays an important role in the 
coverage of the PTS estimator. The minimisation problem (\ref{eq_PTS}) is not convex since it is equivalent to a 
quadratic mixed  integer programming problem, nevertheless there exists a unique solution under mild assumptions
on the data (see section \ref{Sec3}).

Let us define with $f_{PTS}(\bb,k)$ the objective function of  (\ref{eq_PTS}).
We have the following straightforward observations regarding the optimisation problem in (\ref{eq_PTS}), where the set of 
feasible solutions is all $(\bb,k)$ for $\bb\in \mathbb{R}^{p}$ and $k=0,\ldots,n$: 
\begin{itemize}
\item[i)] For any feasible solution $(\bb,k)$ there exists an associated set of $k$ observations as determined by the
    ordering
\[
r(\bb)_{(1)}^{2} \le r(\bb)_{(2)}^{2} \le \cdots \le r(\bb)_{(k)}^{2} \le \cdots \le  r(\bb)_{(n)}^{2}. 
\]
\item[ii)] For any $\bb\in \mathbb{R}^{p}$ there exists $k^{*}$ such that 
\[
f_{PTS}(\bb,k^{*}) \leq f_{PTS}(\bb,k), \;\;\forall k=0,\ldots,n, 
\]
where $k^{*}$ is determined by those observations with squared residuals w.r.t. $\bb$ less than their respective
penalties. Therefore at optimality, $\bb\in \mathbb{R}^{p}$  determines $k$ uniquely. 
\item[iii)] Since the second term of the objective function in (\ref{eq_PTS}) is independent of $\bb$, for any feasible solution $(\bb,k)$ 
\[
f_{PTS}(\bb_{OLS},k) \leq f_{PTS}(\bb,k),
\]
where $\bb_{OLS}$ is the ordinary least squares estimate obtained for the set of observations so defined by
$(\bb,k)$. 
\end{itemize}
We summarise the above in the following lemma which will be used later. 
\begin{lemma}\label{lemma_pts_ols}
If $(\bb_{PTS}, k_{PTS})$ is the solution obtained by (\ref{eq_PTS}) for sample data $(\vc{X,y})$ then
\[
OLS(\vc{X}_{k_{PTS}},\vc{y}_{k_{PTS}}) = \bb_{PTS},
\]
where $(\vc{X}_{k_{PTS}},\vc{y}_{k_{PTS}} )$ is the submatrix and subvector of $(\vc{X,y})$ respectively, as determined
by the set of $k_{PTS}$ observations defined by $(\bb_{PTS}, k_{PTS})$. 
\end{lemma}

In view of the above lemma and (\ref{eq_adj_residual}) we can now state the {\bf general principle} of the PTS estimator, that is to delete 
an observation $i$ if its squared adjusted residual is larger its the penalty cost 
\begin{equation}\label{eq_20}
\alpha_{i}^2 > (c\hat{\sigma})^{2}.
\end{equation}
Therefore, the adjusted residual has as a threshold the square root of the deleting penalty $c\hat{\sigma}$.
Given that the PTS estimate is the OLS of the clean data set, the resulted adjusted residuals must be smaller or equal than the 
penalties
\begin{equation}\label{eq_21}
\frac{|r_{i}|}{\sqrt{1-h_{i}^{*}}} < c\hat{\sigma}, \;\; 0\leq i\le k.
\end{equation}
where $h_{i}^{*}$ is the leverage of the $i^{th}$ observation in the $k$ observations. Otherwise, the 
$i^{th}$ observation can not remain in the coverage sample due to the PTS principal.

The PTS as defined in (\ref{eq_PTS}) ``trims'' outliers from the data but instead of discarding a fixed number of 
observations, a fixed threshold $(c\hat{\sigma})$ for the allowable size of the adjusted residuals is used.

\subsection{Properties of the PTS}
Most of the robust properties of the PTS estimator are inherited by the use of the robust scale $\hat{\sigma}$, which in our
case will be  obtained from the LTS estimator with coverage $(n+p+1)/2$. In what follows 
we will employ the notation used in~\cite{RouLer:87}.

Let $T(\vc{X},\vc{y})$ be an estimator
for some sample data $(\vc{X,y})$. 
An estimator $T$ is called {\em regression equivariant} if 
\[
T(\vc{X}, \vc{y}+\vc{Xv}) = T(\vc{X,y}) + \vc{v}, \;\;\;\mbox{for any}\;\;\vc{v}\in\mathbb{R}^{n}, 
\]
{\em scale equivariant} if 
\[
T(\vc{X},c\vc{y}) = c T(\vc{X,y}), \;\;\;\mbox{for any}\;\;c\in\mathbb{R}, 
\]
and {\em affine equivariant} if 
\[
T(\vc{XA,y}) = \vc{A}^{-1} T(\vc{X,y}), \;\;\;\mbox{for any nonsingular}\;\;\vc{A}\in\mathbb{R}^{p\times p}.
\]

\begin{lemma}\label{lem_equivariance} 
The PTS estimator is regression, scale and affine equivariant. 
\end{lemma}
\begin{proof}
For regression equivariance, by Lemma~\ref{lemma_pts_ols} there will exist some subset of $k$ observations were 
\begin{eqnarray*}
PTS(\vc{X}, \vc{y}+\vc{Xv}, \vc{p}) & = & (\vc{X}_{k}^{T}\vc{X}_{k})^{-1}\vc{X}_{k}^{T}(\vc{y}_{k} + \vc{X}_{k}\vc{v}) \\
                                    & = & (\vc{X}_{k}^{T}\vc{X}_{k})^{-1}\vc{X}_{k}^{T}\vc{y}_{k} + \vc{v} \\
                                    & = & PTS(\vc{X}, \vc{y}, \vc{p}) + \vc{v}.
\end{eqnarray*}
Similarly for scale and affine equivariance. 
\end{proof}

The PTS estimator determines that subset of observations whose deletion produces the largest reduction in the 
sum of squared residuals. The LTS estimator determines the subset of size $k$ with the minimum sum of squared residuals. 
The relationship between the two estimators can be seen by the following propositions. 

\begin{proposition}\label{Pro1}
If the PTS estimator for sample data $(\vc{X,y})$ converges to the solution $(\bb_{PTS},k_{PTS})$, then 
$LTS(\vc{X,y},k_{PTS}) = \bb_{PTS}$. 
\end{proposition}
\begin{proof}
Observe that any $k$ observations will have the same sum of penalties, therefore we will
have
\begin{eqnarray*}
LTS(\vc{X,y},k_{PTS})& = &  \arg\min_{\bb}  \sum_{i=1}^{k_{PTS}}  r(\bb)_{(i)}^2 \\ 
                   & = &  \arg\min_{\bb} \left( \sum_{i=1}^{k_{PTS}}  r(\bb)_{(i)}^2  + (n-k_{PTS}) c\hat{\sigma} \right) \\ 
                   & = & \bb_{PTS}
\end{eqnarray*}
\end{proof}

Conversely under mild assumptions, from the LTS estimator given in (\ref{eq_LTS}) and using the largest residual $r_{k}$ as the deleting penalty, 
the PTS estimator given in (\ref{eq_PTS}) leads to the same results.
\begin{proposition}\label{Pro2}
If for sample data $(\vc{X,y})$ and coverage $k$ we have $LTS(\vc{X,y},k)=\bb_{LTS}$ while
the largest residual $r(\bb_{LTS})_{k}$ is an increasing function with respect to $k$,
then for fixed penalty $p_{i}=(r(\bb_{LTS})_{k})^2,\;\;\; i=1,\ldots,n,$ we have 
$PTS(\vc{X,y,p})=\bb_{LTS}$. 
\end{proposition}
\begin{proof}
Let $PTS(\vc{X,y,p})=(\bb_{PTS},k_{PTS})$.
By Proposition~\ref{Pro1} enough to show that $k_{PTS}=k$. Moreover the largest residual $r(\bb_{PTS})_{k_{PTS}}$ will also 
be an increasing function with respect to $k_{PTS}$.  
We will have for any  $k \geq k_{PTS}$
\begin{eqnarray*}
\sum_{i=1}^{k_{PTS}} r_{i}^{2} + \sum_{i=k_{PTS}}^{n} p_{i} & = & \sum_{i=1}^{k}r_{i}^{2} - \sum_{i=k_{PTS}}^{k} r_{i}^{2} + \sum_{i=k}^{n}p_{i}+ \sum_{i=k_{PTS}}^{k} p_{i} \\
& = & \sum_{i=1}^{k}r_{i}^{2} + \sum_{i=k}^{n}p_{i} +  \sum_{i=k_{PTS}}^{k} (p_{i} - r_{i}^{2} ) \\
\end{eqnarray*} 
where the terms in the last summation are all nonnegative since $p_{i} \geq (r(\bb_{LTS})_{i})^2,$ $i=k_{PTS},\ldots,k$. 
Therefore the minimum is obtained for $k=k_{PTS}$. Similarly for $k\leq k_{PTS}$. 
\end{proof}

In the LTS estimator the coverage $k$ has to be chosen between $n/2$ and $n$. For coverage $k=[(n+p+1)/2]$ the maximum 
breakdown point is obtained by loosing efficiency whereas for $k=n$ the obtained breakdown value is reduced to $1/n$. 
As it is the case with most robust estimators, there is a trade-off between robustness and efficiency. 
The advantage of the PTS procedure is that there is no need to define a priory the coverage $k$. For given penalty 
$(c\hat{\sigma})^2$, where $\hat{\sigma}$ is a high breakdown robust scale, the PTS can resist against all influential 
observations with large adjusted residuals (i.e.  $\alpha_{i} > c \hat{\sigma}$).

\subsubsection{Exact Fit Property and Breakdown Point}\label{sssec_exact}
In this section we will investigate two common robust properties for the PTS estimator, namely the exact fit and breakdown
points. 
The {\em exact fit point} of an estimator $T$ is defined as the minimum fraction of observations on a hyperplane, necessary to 
guarantee that the estimator coincides with that hyperplane~\cite{YohZam:88}. Formally
\begin{eqnarray}\label{eq_exact_fit_point}
\delta^{*}(T,\vc{X,y}) & := & \min \left\{ \frac{m}{n} : \exists \bb \;\;\mbox{such that}\;\; \vc{X}_{m}\bb = \vc{y}_{m}, T(\vc{X,y}) = \bb \right\},
\end{eqnarray}
where $(\vc{X}_{m},\vc{y}_{m})$ is the submatrix and subvector of $(\vc{X},\vc{y})$ respectively,  as determined by the set of $m$ observations. 
A dual notion is used in~\cite{RouLer:87}, where the exact fit point is taken as the minimum fraction of observations {\em not}
on a hyperplane that will force the estimator to give a different estimate from that hyperplane. 
An estimator is said to have the {\em exact fit property} when  $\delta^{*} \leq 0.5$. 

The {\em breakdown point} of an estimator $T$ is defined as the minimum fraction of observations that could take arbitrary 
values and force the estimator to vary indefinitely from its original estimate. This is the finite version of breakdown
point as it was introduced in~\cite{DonHub:83}, and could be stated formally as 
\begin{eqnarray}\label{eq_breakdown_point}
\epsilon^{*}(T,\vc{X,y}) & := &  \min \left\{ \frac{m}{n} : \sup_{\vc{X}_{m}, \vc{y}_{m}} \| T(\vc{X,y}) - T(\vc{X}_{m},\vc{y}_{m}) \| = \infty \right\},
\end{eqnarray}
where $(\vc{X}_{m},\vc{y}_{m})$ is the submatrix and subvector of $(\vc{X},\vc{y})$ respectively,  as determined by the set of $m$ observations. 
The relationship between exact fit point and breakdown point for regression equivariant estimators is 
\begin{equation}\label{eq_exact_breakdown}
\delta^{*} \geq 1 - \epsilon^{*}
\end{equation}
where under mild assumptions equality also holds.

\begin{proposition}\label{pro_exact_fit}
If for sample data $(\vc{X,y})$ there exists some $\bb$ such that strictly more than $\frac{n+p-1}{2}$ of the observations satisfy
$\vc{x}_{i}\bb = y_{i}$ and are in general position, then the PTS solution with robust scale $\sigma_{LTS}$ equals $\bb$. 
\end{proposition}
\begin{proof}
Since the above is true for the LTS estimator (see corollary in page 134 of~\cite{RouLer:87}), we will have $\sigma_{LTS}=0$. Therefore
the penalties as defined in (\ref{eq_PTS}) will be $\vc{p}=(\epsilon,\ldots,\epsilon)$ and at optimality it will hold
\[
PTS(\vc{X,y,p}) = LTS(\vc{X,y},k_{LTS})= \bb,
\]
for $k_{LTS}= \frac{n+p+1}{2}$. 
\end{proof}
It is known~\cite{RouLer:87} that for regression and scale equivariant estimators it holds
\[
\epsilon^{*} \leq \frac{n-p+2}{2n}.
\]
Below we show that the PTS estimator has a high breakdown point. 
\begin{proposition}\label{pro_breakdown}
The breakdown point of the PTS estimator with robust scale $\sigma_{LTS}$ is bounded from below as
\[
\epsilon_{PTS}^{*} \geq \frac{n-p-1}{2n}.
\]
\end{proposition}
\begin{proof}
By Lemma~\ref{lem_equivariance} and (\ref{eq_exact_breakdown}) we know that 
\[
\delta^{*}_{PTS} \geq 1- \epsilon^{*}_{PTS}, 
\]
while by Proposition~\ref{pro_exact_fit} we get 
\[
\delta^{*}_{PTS} \leq \frac{n+p+1}{2n}. 
\]
Therefore by the above
\[
\epsilon^{*}_{PTS} \geq 1 - \frac{n+p+1}{2n} = \frac{n+p-1}{2n}. 
\]
\end{proof}

\subsubsection{Efficiency}

Let $(\vc{X,y})$ be sample data where the random errors $\vc{u}$ as given in the model (\ref{eq_1}) follow the
normal distribution, $\vc{u} \sim  N(0,\sigma^{2})$. 

Consider now that $n\rightarrow \infty$. We know from~\cite{RouLer:87} that
the variance $\sigma_{LTS}$ obtained from the LTS estimator is consistent (i.e. $\lim_{n\rightarrow \infty} \sigma_{LTS}=\sigma$).
Moreover for each observation $i$ we have that the leverage $h_{i}=0$, which means that by the PTS general principle an
observation $i$ will be deleted if
\[
a_{i}^{2} = r(\bb_{OLS})_{i}^{2} > c \hat{\sigma}^{2} 
\]
Therefore, under Gaussian conditions on the random errors and for large $n$, we can say that the PTS estimator with
robust scale $\hat{\sigma} = \sigma_{LTS}$ and cut-off parameter $c\in [2,3]$, deletes a very small fraction of 
observations as outliers.

\subsection{Penalties Computation}\label{subsec_penalties}

\subsubsection{Robust scale estimate $\hat{\sigma}$}

The key to the success of the PTS estimator is to use a robust scale $\hat{\sigma}$ for evaluating the proper 
penalties. This robust scale $\hat{\sigma}$ is obtained via the LTS estimator with coverage $k_{LTS}=[(n+p+1)/2]$ as follows. 
Initially a preliminary error scale $\hat{s}$ is computed 
\[
\hat{s} = c_{k_{LTS},n} \sqrt{ \frac{1}{k_{LTS}} \sum_{i=1}^{k_{LTS}} r(\bb_{LTS})^{2}_{(i)} },
\]
where the constant $c_{k_{LTS},n}$ is so chosen in order to make the scale estimation consistent at the Gaussian model, that is
\[
c_{k_{LTS}, n} = 1 / \sqrt{1 - \frac{2n}{k_{LTS} \alpha_{k_{LTS},n} } \Phi\left(\frac{1}{\alpha_{k_{LTS},n}}\right) },
\]
where
\[
\alpha_{k_{LTS},n} = \frac{1}{\Phi^{-1}(\frac{k_{LTS}+n}{2n})}. 
\]
Then we improve the efficiency results of the LTS by applying a re weighted procedure (see~\cite{RouLer:87}). First the 
standarized residuals  $r(\bb_{LTS})_{i} / \hat{s}$ are computed, and they are used to determine a weight for each
observation as follows: 
\[
w_{i} = 
\left\{ 
\begin{array}{rl}
1 & \textrm{if  $\frac{r(\bb_{LTS})_{i}}{\hat{s}} |\hat{s}| \leq 2.5$,} \\
0  & \textrm{otherwise.}
\end{array} \right.
\]
The final robust scale is then given by
\begin{equation}\label{eq_error_scale}
\hat{\sigma} = \sqrt{\frac{ \sum_{i=1}^{n} w_{i} r(\bb_{LTS})_{i}^{2} }{\sum_{i=1}^{n} w_{i} - p} }.
\end{equation}

\subsubsection{Unmasking Multiple High Leverage Outliers}

Based on the PTS principal, an observation is deleted if its adjusted residual is greater than the square root of the 
penalty cost $\alpha_{i} \ge c\hat{\sigma}$. For $y$-outliers and even for some $\vc{x}$-outliers
such an approach has successful performance. 
Unfortunately the masking problem arises when there is a group of high leverage points in the 
same direction. As it is known, the leverage value $h_{i}$ can be distorted by the presence of a collection of 
points, which individually have small leverages but collectively form a high leverage group. Pe\~na and 
Yohai~\cite{PenYoh:99} point out that the individual leverage $h_{i}$ of each point might be small $h_{i}<<1$, whereas 
the final residual  may appear to be very close to $0$, and this is a masking problem. 
Thus in a set of identical high leverage outliers, the adjusted residual is masked and might be too small 
$\alpha_{i} << c_{i}\hat{\sigma}$, and as a consequence a masked high leverage outlier may not be deleted.

In order to overcome the distortion caused by the masking problem and unmask the leverage $h_{i}$ of the potential high 
leverage points, we are based on the Minimum Covariance Determinant (MCD) procedure (Rousseeuw and Van 
Driessen~\cite{RouDrie:99}). For given coverage $k$, the MCD procedure yields the ``clean'' $\vc{X}_{k}$ matrix 
where $n-k$ vectors $\vc{x}_{i}$ corresponding to high leverage observations have been removed from the original 
matrix $\vc{X}$. Thus, for coverage $k$ close to $51\%$ (i.e. $k=[(n+p+1)/2]$), the leverage of each point 
$(\vc{x}_{i},y_{i}), \; i=k+1,\ldots,n$ as it joins the clean data subset taken from MCD is  
\[
h_{i}^{*} = \vc{x}_{i}^{T}(\vc{X}_{k,i}^{T}\vc{X}_{k,i})^{-1}\vc{x}_{i}, \;\; \text{for} \;\; i=k+1,\ldots,n,
\]
where $\vc{X}_{k,i}$ is the matrix $\vc{X}_{k}$ plus the row $\vc{x}_{i}$. 
In the above $h_{i}^{*}$ is the unmasked leverage of each one of the potential high leverage points
which can be considered as the leverage at the clean data set yielded from the MCD procedure. For the rest of the 
observations $(\vc{x}_{i},y_{i})$ the new leverages now are 
\[
h_{i}^{*} = \vc{x}_{i}^{T} (\vc{X}_{k}^{T}\vc{X}_{k})^{-1}\vc{x}_{i}, \;\; \text{for} \;\; i=1,\ldots,k.
\]
Given that the high leverage identical outliers have been removed from the clean matrix $\vc{X}_{k}$ due to the MCD 
procedure, the new leverage values $h_{i}^{*}$ of the masked points will appear larger than the original $h_{i}$.
Thus, in order to overcome the masking problem we transform the adjusted residuals as follows:
\begin{equation}\label{eq_40}
\alpha_{i}^{*}=\frac{r_{i}}{\sqrt{1-h_{i}}}  \cdot \frac{\sqrt{1-h_{i}}}{\sqrt{1-h_{i}^{*}}} = \frac{r_{i}}{\sqrt{1-h_{i}^{*}}}.
\end{equation} 
Based on the reliable results of the MCD procedure, the new robust adjusted residual in (\ref{eq_40}) is a good 
diagnostic criterion for all observations even for the multiple high leverage outliers.
Incorporating the new robust adjusted residual now in the PTS estimator, we can see how the penalties get individualised
for each observation since an observation now will be considered as an outlier if 
\[
\alpha_{i}^{*} > (c\hat{\sigma}) \Rightarrow r_{i} > c \sqrt{1-h_{i}^{*}} \hat{\sigma},
\]
that is the penalty for each observation is defined as 
\begin{equation}\label{eq_penalties}
p_{i} := (c \sqrt{1-h_{i}^{*}} \hat{\sigma})^{2}
\end{equation}
where $c$ is the cut-off parameter, $\hat{\sigma}$ the robust scale while $h_{i}^{*}$ the robust leverage.

Thus, the PTS estimator tends to remove the masked high leverage observations from the data set, unless their residuals are very small.
However, according to the above analysis, good high leverage points may also appear as extreme 
points with large leverages $h_{i}^{*}$ and the yielded adjusted residual threshold may be too small. As a consequence, 
some of the good high leverage points may be deleted from the data set. This affects only the efficiency of the solution.
However, the efficiency of the final estimate is improved by testing each potential outlier one by 
one in the reinclusion stage of the PTS procedure which is described in the next section.

\subsection{Reinclusion of Observations}

In order to improve the efficiency of the $\hat{\sigma}_{k}$ all observations with small studentized residuals 
are reincluded into the clean subset of size $k$. We follow the reinclusion test for each of the $n-k$ observations, similarly 
to Hadi and Simonoff~\cite{HadSim:93} and Pena and Yohai~\cite{PenYoh:99}. All the $n-k$ points deleted initially 
stage are tested one by one using the studentized predicted error for outlyingness
\[
t_{i}=\frac{y_{i}-\vc{x}_{i}^{T}\bb_{PTS}}{\hat{\sigma}\sqrt{1+h_{i}}},\;\; k+1\le i \le n,
\]
where $h_{i}=\vc{x}_{i}^{T}[\vc{X}_{k}^{T}\vc{X}_{k}]^{-1}\vc{x}_{i}$. 
Under normality assumptions, $t_{i}$ follows Student's $t$ distribution with $k$ degrees of freedom. Therefore, 
every observation $(\vc{x}_{i},y_{i})$,  is reincluded in the subset of the clean data if 
$t_{i}\le t_{\alpha/2,k}$. We have found empirically that for small samples, $t_{\alpha/2,k}=2$ works well.
The final PTS estimate is the OLS solution of the resulting data set after the reinclusion.


\subsection{Quadratic Programming Formulation}\label{Sec3}
The PTS estimator as defined in (\ref{eq_PTS}) can be stated as a Quadratic Mixed Integer Programming (QMIP)
problem, by expanding the residuals and adding 0-1 decision variables, as follows: 
\begin{eqnarray}
\label{eq_pts_qmip}
\min_{\bs{\beta}} & & \sum_{i=1}^{n} \left( r_{i}^{2}+\delta_{i}(c\hat{\sigma})^2\right) \nonumber\\
\mbox{s.t.} & & \vc{x}_{i}^{T}\bs{\beta}+r_{i} \geq y_{i}-s_{i}\\
                                              & & \vc{x}_{i}^{T}\bs{\beta}-r_{i} \leq y_{i}+s_{i}\nonumber\\
                                              & & s_{i} \leq \delta_{i} M\nonumber\\
                                              & & \delta_{i}\in \{0,1\} \nonumber\\
                                              & & r_{i},s_{i} \geq 0 \;\;\; i=1,\ldots, n, \nonumber
\end{eqnarray}
where $s_{i}$ can be regarded as  the pulling distance for moving an outlier towards the regression line, 
$\delta_{i}$ is a decision variable which takes the value of 1 if observation $i$ is to be removed from the clean
data and 0 otherwise, and $M$ is an upper bound on the residuals $r_{i},i=1,\ldots,n$. 
Given any fixed $\bs{\delta}\in\{0,1\}^{n}$, from the $2^{n}$ possible ones, and using matrix notation problem
(\ref{eq_pts_qmip}) becomes
\begin{eqnarray}
\min_{\bs{\beta}} & & \vc{r}^{T}\vc{r}+\bs{\delta}^{T}\vc{p} \nonumber\\
\mbox{s.t.} & & \vc{X}\bs{\beta}+\vc{r} \geq \vc{y}-\vc{s}\nonumber\\
                                              & & \vc{X}\bs{\beta}-\vc{r} \leq \vc{y}+\vc{s}\nonumber\\
                                              & & \vc{s} \leq \bs{\delta}M\nonumber\\
                                              & & \vc{r},\vc{s} \geq \vc{0},\nonumber
\end{eqnarray}
where $\vc{p}=((c\hat{\sigma})^{2},(c\hat{\sigma})^{2},\ldots,(c\hat{\sigma})^{2})^{T}$, $\vc{r}=(r_{1},\ldots,r_{n})^{T}$, 
$\vc{s}=(s_{1},\ldots,s_{n})^{T}$, $\vc{y}=(y_{1},\ldots,y_{n})^{T}$, $\bs{\beta}=(\beta_{1},\ldots,\beta_{n})^{T}$ and 
the matrix $\vc{X}=[\vc{x}_{1},\vc{x}_{2},\ldots,\vc{x}_{n}]^{T}$. This problem has linear constraints and a convex 
quadratic objective function, since the Hessian of $\vc{r}^{T}\vc{r}$ has nonnegative eigenvalues (and it is therefore 
positive semi-definite). Therefore we have a convex program, which will have a unique global optimum solution according 
to the Karush-Kuhn-Tucker optimality conditions~\cite{KaKuTu:93}. Considering that there is a finite number of possible 
$\bs{\delta}$, we can conclude that under the assumption that the data are in general position, problem (\ref{eq_pts_qmip})
or equivalently (\ref{eq_PTS}) has a unique solution. 

Due to the high computational complexity of the resulting QMIP problem the PTS estimator as defined by
(\ref{eq_pts_qmip}) cannot be solved exactly in reasonable time, even for moderately sized problem instances (i.e. $n<50$). This 
is because the computational time required by even the state of the art integer solvers such as CPLEX, grows 
exponentially with respect to the size of the problem instance. Moreover it is almost certain (unless $P=NP$) that 
there does not exist an exact  polynomial time algorithm that solves (\ref{eq_pts_qmip}). Therefore the need arises for a
specialised approximation algorithm, which we present in the next section.

\section{Fast-PTS}\label{sec_fast-pts}

Let us provide a combinatorial formulation of problem~(\ref{eq_pts_qmip}) in order to facilitate 
the discussion of the algorithm. Observe that 
$\delta_{i} =1$ implies $r_{i}=0$,  while for $\delta_{i}=0$ we have that
$r_{i} = r(\bs \beta)_{i}$ which is the residual error of observation $i$ with
respect to the regression vector ${\bs \beta}$. If we let the set of observations
be $O$ we could equivalently formulate problem~(\ref{eq_pts_qmip}) as follows:
\begin{eqnarray} 
\min_{T\subseteq O} & &  L(T):=\sum_{i\in T} r({\bs\beta}_{T})_{i}^{2} + \sum_{i\in O/T} p_{i} \label{eq_pts_comb}
\end{eqnarray}
where $p_{i}$ is the associated penalty for
observation $i$ defined in (\ref{eq_penalties}), and ${\bs\beta}_{T}$ is the OLS solution for the set $T$ of
observations uniquely defined as
\begin{equation}\label{eq_normal}
{\bs\beta}_{T} := (\vc{X}_{T}^{T}\vc{X}_{T})^{-1}\vc{X}_{T}^{T}\vc{y}_{T},
\end{equation}
where $\vc{X}_{T}, \vc{y}_{T} $ are submatrices of $\vc{X}$ and $\vc{y}$
respectively, whose rows are  indexed by the observations in $T$. The equivalence of
problem (\ref{eq_pts_comb}) with (\ref{eq_pts_qmip}) stems from the one-to-one correspondence
between ${\bs\beta}_{T}$ and $T\subseteq O$, as it is defined by the normal equations in (\ref{eq_normal}). 
The following necessary conditions for optimality can be stated. 
\begin{proposition}\label{pro_fast-pts}
If $T\subseteq O$ is an optimal solution to (\ref{eq_pts_comb}) then 
\begin{eqnarray*}
i) & & r(\bb_{T})_{i}^{2} \leq p_{i}, \;\; \forall i\in T \\
ii)& &  r(\bb_{T})_{i}^{2} \geq p_{i}, \;\; \forall i\in O/T
\end{eqnarray*}
\end{proposition}
\begin{proof}
For the case {\em i)} let us define the set
\[
S_{1} := \{ i\in T : r(\bb_{T})_{i}^{2} >  p_{i}\}
\]
and assume that $S_{1}\neq \emptyset$. Then 
\begin{eqnarray*}
L(T)=\sum_{i\in T} r(\bb_{T})_{i}^{2}+\sum_{i\in O/T} p_{i} & > & \sum_{i\in T/S_{1}} r(\bb_{T})_{i}^{2}+\sum_{i\in O/\{T/S_{1}\}} p_{i} \\
                             & \geq & \sum_{i\in T/S_{1}} r(\bb_{T/S_{1}})_{i}^{2} + \sum_{i\in O/\{T/S_{1}\}  } p_{i}
\end{eqnarray*}
therefore we have found a set $T/S_{1}$ were $L(T) > L(T/S_{1})$ which is a contradiction to the hypothesis. 
For the case {\em ii)} analogously define the set 
\[
S_{2} := \{ i\in O/T : r(\bb_{T})_{i}^{2} <  p_{i}\}
\]
and assume that $S_{2}\neq \emptyset$. Then
\begin{eqnarray*}
L(T)=\sum_{i\in T} r(\bb_{T})_{i}^{2}+\sum_{i\in O/T} p_{i} & > & \sum_{i\in T} r(\bb_{T})_{i}^{2}+ \sum_{i\in S_{2}} r(\bb_{T})_{i}^{2}   
                                                                \sum_{i\in O/ \{T\cup S_{2}\}} p_{i} \\
                             & \geq & \sum_{i\in T\cup S_{2}} r(\bb_{T\cup S_{2}})_{i}^{2} + \sum_{i\in O/ \{T\cup S_{2}\}} p_{i}
\end{eqnarray*} 
which implies that  $L(T) > L(T\cup S_{2})$ and again it contradicts the original hypothesis of the optimality of $T$. 
Therefore $S_{1}=S_{2}=\emptyset$ and the proposition is proved. 
\end{proof}
It can be easily verified by a counterexample that the above conditions are not sufficient. Nevertheless the number of local optima
which they characterise seems to be much smaller than that characterised by more classic $k$-exchange neighbourhoods which
grow exponentially with respect to the size of the associated QMIP. Specifically, if we define the {\em $k$-exchange neighbourhood} of 
a feasible solution $T$ to (\ref{eq_pts_comb}) as
\begin{equation*}
N(T)_{k} := \{ S\subseteq O: |T\Delta S| \leq k \}
\end{equation*}
where $\Delta$ stands for the symmetric difference of sets, then we have the trivial necessary conditions for optimality 
\begin{equation}\label{eq-kxnge}
T \;\;\mbox{optimal to (\ref{eq_pts_comb})} \;\;\Rightarrow \;\; L(T) \leq L(S), \;\forall S\in N(T)_{k}.
\end{equation}
A local search algorithm based on (\ref{eq-kxnge}) will have to have to search the $k$ neighbourhood of a feasible solution entirely, 
which will require an exponential number of steps on $k$ with either a breadth-first or a depth-first search strategy. Therefore
even for small $k=1,2$ this type of local search is computationally very expensive. 

The first necessary optimality condition of Proposition~\ref{pro_fast-pts} is used to construct a feasible solution 
in a randomised fashion, while the second is used to obtain the local optimum in the region defined by this solution.

\subsection{The Algorithm}
\begin{figure}
\begin{center}
  \fbox{
  \begin{minipage}[b]{3.6in}
    \noindent
    {\bf procedure} {\tt Fast-PTS}($\vc{X,y,p}, n, p$, {\tt MaxIter, RandomSeed}) \\
    1\ \hspace{.2in}$T_{min}:= \{1,\ldots,n\}$;\\
    2\ \hspace{.2in}{\bf do} $k = 1,\ldots,\;${\tt MaxIter} $\rightarrow$\\
    3\ \hspace{.5in}$T:= $ {\tt Construction($\vc{X,y,p}, n, p$,RandomSeed)};\\
    4\ \hspace{.5in}$T:= $ {\tt LocalSearch($\vc{X,y,p}, n, p,T$)};\\
    5\ \hspace{.5in}{\bf if} $(L(T) < L(T_{min}))$ $\rightarrow$ $T_{min} := T$\\
    6\ \hspace{.2in}{\bf od};\\
    7\ \hspace{.2in}{\bf return}(${\bs\beta}_{T_{min}}$)\\
    {\bf end} {\tt Fast-PTS};
  \end{minipage}
  }
\end{center}
\caption{The Fast-PTS algorithm}
\label{fig_fast-pts}
\end{figure}
The Fast-PTS algorithm is presented in pseudocode in Figure~\ref{fig_fast-pts}. It accepts as inputs the 
problem instance data, and the parameters {\tt MaxIter} and {\tt RandomSeed} which specify the maximum
number of iterations and a seed for the random number generator respectively. Initially the best solution
is $T_{min}$ is set to be all the observations, while this solution is updated in line 5 of Figure~\ref{fig_fast-pts}.
Each iteration of the Fast-PTS algorithm is composed of two phases, a
construction phase where a good solution is built in a greedy randomised fashion starting from an empty
solution, and an improvement phase where the solution from the previous phase
is improved to ensure local optimality. At the end of each iteration an approximate
solution to (\ref{eq_pts_comb}) is provided. The maximum number of iterations is
provided by the user, and the best solution found among all iterations is returned by the algorithm
as the approximation to the optimum solution. This randomised procedure for constructing a solution before 
performing local search, has substantial experimental evidence of good performance for NP-Hard combinatorial
optimisation problems (see~\cite{PitRes02a}) since it helps the algorithm to examine a wider area of the 
feasible space without getting entrapped in local optima.


The construction phase of the algorithm is shown in Figure~\ref{fig_construction}.
Let us call a partial solution of (\ref{eq_pts_comb}) say $T\subseteq O$ {\em penalty free} if
\[
r(\bb_{T})_{i}^{2} < p_{i}, \;\; \forall i\in T.
\]
In the construction phase of the algorithm, initially a partial solution
$T\subset O$ is constructed by choosing at random $(p+1)$ observations such that
$T$ is penalty free (line 1 in  Figure~\ref{fig_construction}). Note that any 
$p$ observations are penalty free since they have zero residuals i.e. we have an
exact fit. 
Then this solution is enlarged in a greedy randomised
fashion maintaining its penalty free property. Given any partial solution $T$
the following set of possible candidates is defined
\[
C(T) := \{ j\in O/T : r(\bb_{T\cup j})_{i}^{2} < p_{i}, \;\;\forall i\in T\cup j \}
\]
The elements of $C(T)$ are then sorted in ascending order according to their objective
function values as given in (\ref{eq_21}), that is
\[
L(T\cup j_{1}) \leq L(T\cup j_{2}) \leq \cdots \leq L(T\cup j_{|C(T)|}),
\]
and one observation $s$ is chosen randomly among the first $\max\{1, \alpha |C(T)|\}$ candidates,
and the partial solution is updated (i.e. $T := T \cup \{s\}$).
The parameter $\alpha \in [0,1]$ controls the degree of greediness versus randomness
in the construction of the solution. If $\alpha=1$ the the algorithm constructs the
solution in a randomised fashion while if $\alpha=0$ the solution is purely greedy.
This procedure is repeated until we reach a partial solution $T$ with $C(T) = \emptyset$,
where the construction phase ends.

\begin{figure}
\begin{center}
  \fbox{
  \begin{minipage}[b]{4.2in}
    \noindent
    {\bf procedure} {\tt Construction}($\vc{X,y,p}, n, p$, {\tt RandomSeed}) \\
    1\ \hspace{.2in}$T_{c}:=$ {\tt ConstructRandom}($\vc{X,y,p}, n, p$, {\tt RandomSeed});\\
    2\ \hspace{.2in}{\bf while} ($C(T_{c}) \neq \emptyset)$ $\rightarrow$ \\
    3\ \hspace{.5in}$C(T_{c}) := \emptyset$;\\
    4\ \hspace{.5in}{\bf do} $j \in O/T_{c}$ $\rightarrow$ \\
    5\ \hspace{.8in}{\bf if} ($r(\bb_{T_{c} \cup j})_{i}^{2} < p_{i}, \;\;\forall i\in T_{c}\cup j $) $\rightarrow$ \\
    6\ \hspace{1.1in} $C(T_{c}) := C(T_{c})\cup j$; \\
    7\ \hspace{1.1in}$L(T_{c}\cup j) := \sum_{i\in T_{c}\cup j} r(\bb_{T_{c}\cup j})_{i}^{2} + \sum_{i\in O/T_{c}\cup j} p_{i}$;\\
    8\ \hspace{1.1in}{\tt  inheap}($L(T_{c}\cup j), j$); \\
    9\ \hspace{.8in}{\bf endif}; \\
   10\ \hspace{.5in}{\bf od;} \\
   11\ \hspace{.5in}$t=${\tt random}$[1,\alpha |C(T_{c})|]$; \\
   12\ \hspace{.5in}{\bf do} $j=1,\ldots,t$   $\rightarrow$ \\
   13\ \hspace{.8in} $j_{c}:=$ {\tt outheap()}; \\
   14\ \hspace{.5in}{\bf od;} \\
   15\ \hspace{.5in}$T_{c}:= T_{c}\cup j_{c}$ ;\\
   16\ \hspace{.2in}{\bf endwhile};\\
   17\ \hspace{.2in}{\bf return}($T_{c}$)\\
    {\bf end} {\tt Construction};
  \end{minipage}
  }
\end{center}
\caption{The {\tt Construction} procedure of the Fast-PTS algorithm}
\label{fig_construction}
\end{figure}
In lines 2-16 of the {\tt Construction} procedure shown in Figure~\ref{fig_construction} the main iterations upon
which a solution $T_{c}$ is constructed, are shown. The iterations in lines 4-10 compute the set $C(T_{c})$ of candidate 
elements to be added in the solution, where the penalty free property is maintained in line 5. The cost as well as the
index of the associated candidate element are added into a heap in line 8, which will be later used for sorting these
values. Finally in lines 11-15, an element $j_{c}$ is chosen at random among the best candidates, and added into our current
solution. The procedure {\tt outheap()} returns the index $j$ of the observation with the smallest $L(T_{c}\cup j)$ value
as computed in and inserted into the heap in lines 7 and 8.


The solution $T_{c}$ computed by the construction phase is penalty free, and we also know that
there is no $j\in O/T_{c}$ such that $T_{c}\cup \{j\}$ is penalty free with respect to $\bb_{T_{c}\cup \{j\}}$.
However there may be an observation $j\in O/T_{c}$ such that its residual with respect to $\bb_{T_{c}}$
is less than its respective penalty. That is, the necessary optimality conditions stated in 
Proposition \ref{pro_fast-pts} may not be satisfied. 
Therefore further improvement of the solution provided by the construction phase could be achieved. This
is performed in the improvement phase of the Fast-PTS algorithm, where the solution by the
construction phase is changed to satisfy both necessary optimality conditions. Given a solution
$T_1$ and its associated regressor vector $\bb_{T_1}$, a new solution $T_{2}$ is computed as
\[
T_{2} := \{ i\in O : r(\bb_{T_{1}})_{i}^{2} < p_{i} \},
\]
and this is repeated until $T_{1} = T_{2}$. As it is shown in the proof of Proposition ~\ref{pro_fast-pts}, 
at each iteration $L(T_{1}) \geq L(T_{2})$
therefore the procedure will terminate in a finite number of steps (usually the convergence is in the order
of 4 to 5 steps). This simple local search procedure is depicted in Figure~\ref{fig_local_search}. 
\begin{figure}
\begin{center}
  \fbox{
  \begin{minipage}[b]{2.5in}
    \noindent
    {\bf procedure} {\tt LocalSearch}($\vc{X,y,p}, n, p,T$) \\
    1\ \hspace{.2in}$T_{1}:= T$;\\
    2\ \hspace{.2in}{\bf while}($T_{2} \neq T_{1}$)  $\rightarrow$\\
    3\ \hspace{.5in}$T_{1}:=T_{2}$; \\
    4\ \hspace{.5in}$T_{2}:= \emptyset $;\\
    5\ \hspace{.5in}{\bf do} $i=1,\ldots,n$ $\rightarrow$\\
    6\ \hspace{.8in}{\bf if} ($r(\bb_{T_{1}})_{i}^{2} < p_{i}$) $\rightarrow$\\
    7\ \hspace{1.1in}$T_{2}:= T_{2}\cup i$; \\
    8\ \hspace{.8in}{\bf endif}; \\
    9\ \hspace{.5in}{\bf od}; \\
   10\ \hspace{.2in}{\bf endwhile}; \\
   11\ \hspace{.2in}{\bf return}($T_{1}$);\\
    {\bf end} {\tt LocalSearch};
  \end{minipage}
  }
\end{center}
\caption{The {\tt LocalSearch} procedure of the Fast-PTS algorithm}
\label{fig_local_search}
\end{figure}

Overall the computational complexity of the Fast-PTS algorithm is polynomial on the size of the problem, 
since for a fixed number of iterations, each iteration requires a polynomial number of solutions to an
OLS problem, which is itself polynomially solvable.


\section{Computational Results}\label{sec_comp}

In this section we present the results of the computational experiments performed in order to verify the 
performance of the Fast-PTS algorithm, and the overall robust behaviour of the proposed estimator. First we 
compare the solutions found by the Fast-PTS algorithm for the QMIP problem (\ref{eq_pts_qmip}) with the 
exact solutions found by ILOG's CPLEX software, on a set of artificial problem instances. Then we compare the
robust behaviour of the algorithm with two of the most succesfull algorithms in 
robust regression i.e. the Fast-LTS algorithm as implemented by Rousseeuw and Van Driessen~\cite{RouDrie:06} 
and the Fast-S algorithm as presented by Barrera and Yohai~\cite{BarYoh:06}. These comparisons were performed on the 
following sets of instances: 
\begin{itemize}
\item a set of ``benchmark'' data sets to regression outlier problems widely known in the literature, 
\item a set of Monte Carlo generated data sets, proposed  by  Barrera and Yohai in~\cite{BarYoh:06}, 
\end{itemize}

The Fast-PTS algorithm was implemented on Intel's Fortran Compiler version 10.1 for Linux with the optimisation flags
{\tt  -ipo,  -O3,  -no-prec-div,  -static}, and ILOG's CPLEX integer programming
solver version 9.1 was used. All computational experiments where performed on a Intel Core Duo 2.4GHz computer with 2GB of 
memory running openSUSE Linux 10.3\footnote{The code is available upon request at \tt{pitsouli@gen.auth.gr}}.

\subsection{Exact Solutions}\label{subsec_exact}
In Table~\ref{table_0} we can see the performance of the Fast-PTS algorithm with respect to exact solutions for 
8 problem instances. In each instance the problem was solved exactly by the CPLEX mixed integer quadratic programming
solver, and the Fast-PTS obtained the exact solution for all instances. The number of maximum iterations was set to 
100, although in all cases the optimal solution was obtained in fewer iterations. As we can see from Table~\ref{table_0}, 
the computational effort required for the exact solution grows exponentially with respect to the size of the 
instance, while the Fast-PTS algorithm requires a fraction of a second.  Similar comparison would be desirable for
larger size instances such as the ones presented in section~\ref{subsec_monte_carlo}. This  however is infeasible since
the required CPU time would be excessively large for the exact solution, and the size of the branch and cut tree 
generated by CPLEX exceeds the available memory capacity. Specifically for the problem set in Table~\ref{table_0}, for
instances with $n\geq 128$, CPLEX required larger memory for the tree and terminated the solution process. 
\vspace*{5mm}
\begin{table}[htb]
\centering
{\footnotesize
\begin{tabular}{lccccc}\hline
               &     &     &  Exact       &    CPLEX         &  Fast-PTS      \\
      data set & $n$ & $p$ &  solution    & (CPU seconds) &  (CPU seconds)    \\  \hline
      PTS1     &  48 &  2  &  12642.9     & 6.05           &   0.038          \\ 
      PTS2     &  58 &  2  &  25910.7     & 5.68           &   0.048          \\ 
      PTS3     &  68 &  2  &  17658.1     & 146.4          &  0.064           \\ 
      PTS4     &  78 &  2  &  26158.8     & 42.3           &   0.076          \\ 
      PTS5     &  88 &  2  &  37951.5     & 1368           &   0.096          \\ 
      PTS6     &  98 &  2  &  41289.2     & 2256           &   0.092          \\ 
      PTS7     & 108 &  2  &  48519.3     & 32055          &  0.132           \\ 
      PTS8     & 118 &  2  &  52829.8     & 37948          &   0.152          \\ \hline
\end{tabular}
}
\caption{Comparison of Fast-PTS with CPLEX}\label{table_0}
\end{table}
\vspace*{5mm}

\subsection{Benchmark Instances}\label{subsec_benchmark}

Five widely used benchmark instances were examined in this experimental study. The first four are taken from Rousseeuw and 
Leroy~\cite{RouLer:87} and have been studied by many statisticians in robust literature. The last one is an artificial 
example from Hadi and Simonoff~\cite{HadSim:93}. Table~\ref{table_1} gives the corresponding results for the PTS estimator, 
indicating the name of the data set, its dimension and number of observations, the included outliers, the percentage of 
outliers that have been identified and the running times in seconds. The penalties for deleting outliers are 
$(2\sqrt{1-h_{i}^{*}}\cdot\hat{\sigma})^2$, as have been proposed in section~\ref{Sec2}. 
These results are similar to those reported in the literature for other high breakdown estimators and 
they indicate that the proposed method PTS behaves reliably on these test sets.

\textbf{Telephone Data.} The first data set contains the total number of telephone calls made in Belgium during the years 
1950-1973. During the period 1964-1969, (cases 15-20), another recording system was used, hence these observations are 
unusually high while cases 14 and 21 are marginal. The outliers draw the OLS regression line upwards, masking the true 
outliers, while swamping in the clean cases 22-24 as too low. The high breakdown estimators Fast-LTS and Fast-S correctly 
flags the outliers. Also, our PTS estimator correctly identify the true outliers.

\textbf{Hertzsprung-Russell Stars Data.} This data set contains 47 stars in the direction of Cygnus, where $x$ is the 
logarithm of the effective temperature at the surface of a star and $y$ is the logarithm of its light intensity. Four stars, 
called giants, have low temperature with high light intensity. These outliers are cases 11, 20, 30 and 34 which can be 
clearly seen by a scatter plot. The three high breakdown methods identify successfully the outliers.

\textbf{Modified Wood Gravity Data.} This is a real data set with five independent variables. It consists of 20 cases; 
some of them were replaced by Rousseeuw~\cite{Rousseeuw:84} to contaminate the data by few outliers, namely cases 4, 6, 8 
and 19. Once again, all three methods manage to reveal the outliers.

\textbf{Hawkins, Bradu and Kass Data.} The data have been generated by Hawkins et. al~\cite{HawBraKas:84} for illustrating 
the merits of a robust technique. This artificial data set offers the advantage that at least the position of the good or 
bad leverage points is known. The Hawkins, Bradu and Kass data consists of 75 observations in four dimensions. The first 
ten observations form a group of identical bad leverage points, the next four points are good leverage while the remaining 
are good data. The problem in this case is to fit a hyperplane to the observed data. Plotting the regression residuals from 
the model obtained from the standard OLS estimator, the bad high-leverage point data are masked and do not show up from the 
residual plot. Some robust methods not only fail to identify the outliers, but they also swamp in the good cases 11-14. 
The Fast-LTS and Fast-S estimators correctly flag the outliers. For the PTS estimator, the MCD procedure reveals the first 
14 points of this data set as high leverage points. Application of the PTS to these data, starting with robust scale 
estimate about $\hat{\sigma}=0.61$ and downweighting the penalty cost with penalties from (\ref{eq_penalties}), rejects only the 
first 10 points as outliers, which are known as the bad leverage points.

\textbf{Hadi Data.} These data have been created by Hadi and Simonoff~\cite{HadSim:93}, in order to illustrate the 
performance of various robust methods in outlier identification. The two predictors were originally created as uniform 
$(0, 15)$ and were then transformed to have a correlation of $0.5$. The depended variable was then created to be consistent 
with the model $y=x_{1}+x_{2}+u$ with $u \sim N(0, 1)$. The first $3$ cases $(1-3)$ were contaminated to have predictor 
values around $(15, 15)$, and to satisfy $y=x_{1}+x_{2}+4$. Scatterplots or diagnostics have failed to detect the outliers. 
Many identification methods fail to identify the three outliers. Some bounded influence estimates have largest absolute 
residual at the clean case $17$, indicating potential swamping. The efficient high breakdown methods Fast-LTS and Fast-S do 
identify the three outliers as the most outlying cases in the sample, but the residuals are too small to be considered 
significantly outliers. In contrast, robust methods proposed by Hadi and Simonoff~\cite{HadSim:93} and Fast-PTS estimator 
identify correctly the clean set $4-25$, with each of the cases $1-3$ having residuals greater than $3.78$.

\begin{table}[htb]
\centering
{\footnotesize
\begin{tabular}{lcclcc}\hline
                                             &     &     &             &  \%outliers & time         \\
      data set                                 & $n$ & $p$ &  outliers         &  identified & (CPU seconds)    \\ \hline
      Telephone                        &  24 &  2  &15,16,17,18,19,20&  100    &   0.05   \\ 
      Stars          &  47 &  2  &11,20,30,34      &  100    &   0.18   \\ 
      Wood           &  20 &  6  &4,6,8,19         &  100    &         0.06   \\ 
      Hawkins        &  75 &  4  &1,2,3,4,5,6,7,8,9,10&100   &   0.50   \\ 
      Hadi                               &  25 &  3  &1,2,3                                      &  100          &   0.08   \\ \hline
\end{tabular}
}
\caption{Performance of the Fast-PTS algorithm on some small data sets.}\label{table_1}
\end{table}

\subsection{Monte Carlo Simulation}\label{subsec_monte_carlo}

To explore further the properties of the PTS method, we have performed an extensive set of simulation experiments for larger 
sample sizes and observation dimensions taken from Barrera and Yohai~\cite{BarYoh:06}. The experiments compare the 
performance of our Fast-PTS algorithm with the results of Fast-LTS and Fast-S algorithms. 

Barrera and Yohai~\cite{BarYoh:06} considered a model as in (\ref{eq_1}) with $\vc{x}=(1,x_{1},x_{2},\ldots,x_{p})$, where a proportion 
$(1-\epsilon)$ in the data samples $\vc{x}=(x_{1},x_{2},\ldots,x_{p},y)$ follow a multivariate normal distribution. 
Without loss of generality the mean vector is taken equal to 0 and the identity as the covariance matrix. These observations 
follow the model in (\ref{eq_1}) with $\bs{\beta}=0$. The contaminated observations are high leverage outliers with 
$\vc{x}=(1,100,0,\ldots,0)$ and $y=\text{slope}\times 100$, where the slope of contamination takes values from 0.9 to 2.0. 

Under high leverage contaminations, the objective functions of the three high breakdown estimators typically have two 
distinct types of local minimum, that is one close to the true value of the regression parameter and a second one close 
to the slope of the outliers.

\begin{table}[htb]
\centering
{\footnotesize
\begin{tabular}{lcccccccccccc}\hline
&   \multicolumn{12}{c}{slope} \\ 
             method   & 0.9 & 1.0 & 1.1 & 1.2 & 1.3 & 1.4 & 1.5 & 1.6 & 1.7 & 1.8 & 1.9 & 2.0      \\ \hline
           Fast-PTS   & 58  & 44  & 21  & 8   & 3   & 2   & 1   & 1   &  1  &  0  &  0  &  0      \\ 
           Fast-S     & 88  & 60  & 29  &  8  & 1   & 0   & 0   & 0   &  0  &  0  &  0  &  0      \\
           Fast-LTS   & 100 & 100 & 95  & 86  & 67  & 42  & 21  & 9   &  4  &  1  &  0  &  0      \\ \hline
\end{tabular}
}
\caption{Percentage of samples where convergence occured to the wrong local minimum. $n=400$, $p=36$ and $10\%$ of outliers located at $(\vc{x}_{1},\ldots,\vc{x}_{36},y)=(1,100,0,\ldots,0,slope\times 100)$.}\label{table_2}
\end{table}

Two measures of performance are used in this Monte Carlo study to compare the three estimators:
\begin{itemize}
\item The percentage of samples for which each algorithm converged to a wrong local minimum i.e. a local minimum close to 
the slope of the outliers.
\item The mean square error (MSE) of the parameter estimate $\hat{\bs{\beta}}$.
\end{itemize}

Several different values of the slope of contamination were used to determine the behaviour of the three algorithms. 
Its range varies from 0.9 to 2.0 with sample sizes of $n=400$ and $p=36$. The proportion of outliers $\epsilon$ was set 
to $10\%$.

Tables~\ref{table_2} and~\ref{table_3} contain the percentage of convergence to a wrong local minimum and the MSEs 
respectively. From Table~\ref{table_2} we observe that the Fast-PTS has the lowest wrong convergence with slope between 
$0.9$ and $1.1$. For larger slope the results are comparable for all estimators. In Table~\ref{table_3} it is shown that 
the Fast-PTS has the lowest MSE for the contaminated data with slope between $0.9$ and $1.9$. For the rest of the data, the 
MSE of the Fast-PTS is comparable with that of Fast-LTS and Fast-S. The main conclusion from Tables~\ref{table_2} 
and ~\ref{table_3} is that for contamination with small slope, the Fast-PTS perform better than Fast-S and Fast-LTS in 
percentage of wrong convergence. However, for larger slopes, the PTS has important improvement in the MSEs.

\begin{table}[htb]
\centering
{\footnotesize
\begin{tabular}{lcccccccccccc}\hline
&   \multicolumn{12}{c}{slope} \\ 
           method       & 0.9  & 1.0  & 1.1  & 1.2  & 1.3  & 1.4  & 1.5  & 1.6  & 1.7  & 1.8  & 1.9  & 2.0     \\ \hline
           Fast-PTS     & 1.54 & 1.30 & 0.82 & 0.48 & 0.32 & 0.29 & 0.36 & 0.27 & 0.31 & 0.30 & 0.33 & 0.48    \\ 
           Fast-S       & 1.65 & 1.40 & 0.98 & 0.62 & 0.48 & 0.47 & 0.45 & 0.45 & 0.47 & 0.46 & 0.46 & 0.46    \\
           Fast-LTS     & 1.73 & 1.46 & 1.39 & 1.34 & 1.08 & 0.92 & 0.74 & 0.58 & 0.51 & 0.47 & 0.46 & 0.46    \\ \hline 
\end{tabular}
}
\caption{Mean Squared Errors. $n=400$, $p=36$ and $10\%$ of outliers located at 
$(\vc{x}_{1},\ldots,\vc{x}_{36},y)=(1,100,0,\ldots,0,slope\times 100)$.} \label{table_3}
\end{table}

To explore further the comparison of the three estimators, we use again the same data as Barrera and 
Yohai~\cite{BarYoh:06}, with different values of $n$, $p$, considering only one value of the contamination slope, 
$y=100\times 1$ and $\epsilon=10\%$. Tables~\ref{table_4} and ~\ref{table_5} show the average time needed until 
convergence of the algorithm, the percentage of samples where the algorithm converged to a wrong local minimum 
(``$\%$wrong'') and the mean squared error (``MSE'' ) for the three estimators. From Table~\ref{table_4} we observe 
that the Fast-PTS performs significantly better in both performance criteria especially when the data set is of size 
$n=100$ or $n=500$.

\begin{table}[htb]
\centering
{\footnotesize
\begin{tabular}{lc|ccc|cc|cc}\hline
                   & \multicolumn{1}{c}{}    & \multicolumn{3}{c}{Fast-PTS} & \multicolumn{2}{c}{Fast-S} & \multicolumn{2}{c}{Fast-LTS} \\
             $n$   & \multicolumn{1}{c}{$p$} & CPU time & $\%$wrong & \multicolumn{1}{c}{MSE}   & $\%$wrong & \multicolumn{1}{c}{MSE}   & $\%$wrong & MSE   \\ \hline
             100   & 2   &  0.08    & 0.2       & 0.030 & 30.6      & 0.370 &  50.8     & 0.692 \\ 
                   & 3   &  0.09    & 0.2       & 0.052 & 37.2      & 0.506 &  56.0     & 0.890 \\ 
                   & 5   &  0.12    & 0.2       & 0.099 & 46.8      & 0.773 &  64.8     & 1.193 \\ \hline
             500   & 5   &  2.28    & 0.0       & 0.014 & 15.4      & 0.193 &  61.4     & 0.803 \\ 
                   & 10  &  3.78    & 0.0       & 0.029 & 19.8      & 0.292 &  65.8     & 0.935 \\ 
                   & 20  &  8.31    & 0.2       & 0.069 & 30.4      & 0.547 &  79.4     & 1.207 \\ \hline
            1000   & 5   &  8.75    & 0.0       & 0.007 & 7.8       & 0.095 &  61.2     & 0.715 \\ 
                   & 10  & 14.41    & 0.0       & 0.014 & 5.4       & 0.089 &  69.4     & 0.859 \\ 
                   & 20  & 31.11    & 0.0       & 0.028 & 12.4      & 0.208 &  82.4     & 1.158 \\ \hline
\end{tabular}
}
\caption{Samples with $10\%$ of outliers located at $(\vc{x}_{1},\ldots,\vc{x}_{p},y)=(1,100,0,\ldots,0,100)$.}\label{table_4}   
\end{table}

For heavier contamination $\epsilon=0.20$ and slope 2.2, Table~\ref{table_5} shows the results of these simulations. As 
in Table~\ref{table_4} we observe that the Fast-PTS has the best overall performance especially for the small data sets 
$n=100$. However, for larger data sets ($n\ge 500$, the Fast-PTS and Fast-S perform comparable, since the outliers are 
easier to detect by both robust estimators due to the large slope value.

\begin{table}[htb]
\centering
{\footnotesize
\begin{tabular}{lc|ccc|cc|cc}\hline
                   & \multicolumn{1}{c}{}    & \multicolumn{3}{c}{Fast-PTS} & \multicolumn{2}{c}{Fast-S} & \multicolumn{2}{c}{Fast-LTS} \\
             $n$   & \multicolumn{1}{c}{$p$} & CPU time & $\%$wrong & \multicolumn{1}{c}{MSE}   & $\%$wrong & \multicolumn{1}{c}{MSE}   & $\%$wrong & MSE   \\ \hline
             100   & 2  &             0.08 &  0.2              &0.032 &       10.0            &       0.543&  40.9   &2.333  \\ 
                   & 3  &             0.09 &  0.0              &0.056 &       15.6      &     0.947&  51.6   &3.218  \\ 
                   & 5  &             0.11 &  0.8              &0.109 &       34.0            &       2.303&  70.2   &5.241  \\ \hline
             500   & 5  &             2.28 &  0.0              &0.016 &        0.0            &       0.025&  23.4     &1.236  \\ 
                   & 10 &             3.74 &  0.0              &0.032 &        0.0            &       0.076&  33.8     &2.102  \\ 
                   & 20 &             7.82 &  11.2     &0.873 &        9.2            &       0.734&  72.2     &4.276  \\ \hline
            1000   & 5  &             9.40 &  0.0              &0.008 &        0.0            &       0.012&  11.0     &0.542  \\ 
                   & 10 &      14.94 &        0.0              &0.016 &        0.0            &       0.025&  20.4     &1.205  \\ 
                   & 20 &      31.63 &        0.0              &0.030 &        0.0            &       0.055&  49.8     &2.980  \\ \hline
\end{tabular}
}
\caption{Samples with $20\%$ of outliers located at $(\vc{x}_{1},\ldots,\vc{x}_{p},y)=(1,100,0,\ldots,0,220)$.}  \label{table_5}
\end{table}

To continue the implementation of the Fast-PTS algorithm and compare it with the well-known methods discussed in this 
article, we perform extra Monte Carlo experiments to evaluate the performance of our robust procedure. To carry out one 
simulation run, we proceeded as follows. The distributions of independent variables and errors and the values of parameters 
are given. The observations $y_{i}$, were obtained following the regression model as in (\ref{eq_1}) with $p=3$ and $n=50$, 
where the coefficient values are $\beta_{1}=1.20$, $\beta_{2}=-0.80$ and a zero constant term $\beta_{0}=0.0$. We prefer 
the Gauss distribution for the iid error term $u \sim N(0, \sigma^{2}=16^{2})$, while $\vc{x}_{1}$ and ${x}_{2}$ are iid 
values drawn also from normal distributions $N(\mu=20, \sigma^{2}=6^{2})$ and $N(\mu=30, \sigma^{2}=8^{2})$ respectively. 
We consider that the sample may contain three types of outliers, regression outliers (``bad'' high-leverage points), 
``good'' high-leverage points, and response outliers ($y$-outliers).
An extra value is drawn from the uniform distribution $U(a=80, b=220)$ and for the regression outlier is added to 
$x_{1}$ or $x_{2}$, for the ``good'' leverage point is added to $x_{1}$ or $x_{2}$ but the value of the dependent 
variable $y$ follows their contamination, according to the above regression model, for the response outlier is added to $y$.
All simulation results are based on $150$ replications enough to obtain a relative error $<10\%$ with a reasonable 
confidence level of at least $90\%$ for all the simulation estimates. The robust scale estimate $\hat{\sigma}$ from LTS 
with coverage $k=28$ is used throughout the simulation study. 

Tables~\ref{table_6} and ~\ref{table_7} display the results concerning the performance of the three robust estimators 
corresponding to the following cases: Table~\ref{table_6} is based on data contaminated by ``bad'' and ``good'' high 
leverage points whereas Table~\ref{table_7} is based on data contaminated by ``bad'' high leverage outliers (heavier 
contamination).

\begin{table}[htb]
\centering
{\footnotesize
\begin{tabular}{lccc}\hline
                                                &  Fast-PTS  & Fast-S   &  Fast-LTS      \\ \hline
      \%wrong                                   &  2.7    & 13.3    &  16.0                   \\ 
      mean estimate of $\hat{\beta}_{0}$        &  -1.796 & -1.633  & -2.782                   \\ 
      mean estimate of $\hat{\beta}_{1}$        &  1.176  & 1.129   &  1.147                   \\ 
      mean estimate of $\hat{\beta}_{2}$        & -0.766  &-0.758   & -0.732                   \\ 
      mean squared error of $\hat{\beta_{0}}$   &  35.38  & 58.41   &  112.21                    \\ 
      mean squared error of $\hat{\beta_{1}}$   &  0.036  & 0.106   &  0.132                    \\ 
      mean squared error of $\hat{\beta_{2}}$   &  0.015  & 0.030   &  0.049                    \\ 
      computation time (secs)                   &  0.56   &  0.29   &  0.18                   \\ \hline
\end{tabular}
}
\caption{Samples with $32\%$ of outliers ($x$-outliers=6, ``good'' leverage points=4, $y$-outliers=6), $n=50$, $p=3$.
True: $\beta_{0}=0.0$, $\beta_{1}=1.20$, $\beta_{2}=-0.80$.}\label{table_6}
\end{table}

\begin{table}[htb]
\centering
{\footnotesize
\begin{tabular}{lccc}\hline
                                                 &  Fast-PTS  & Fast-S   &  Fast-LTS      \\ \hline
      \%wrong                                    &  26.0      & 61.3     &  50.0                   \\ 
      mean estimate of $\hat{\beta}_{0}$         &  -3.955    & -1.607   & -2.123                   \\ 
      mean estimate of $\hat{\beta}_{1}$         &  1.131     & 0.814    &  0.910                   \\ 
      mean estimate of $\hat{\beta}_{2}$         & -0.646     &-0.569    & -0.600                   \\ 
      mean squared error of $\hat{\beta_{0}}$    &  84.41     & 158.25   &  173.98                    \\ 
      mean squared error of $\hat{\beta_{1}}$    &  0.088     & 0.473    &  0.390                    \\ 
      mean squared error of $\hat{\beta_{2}}$    &  0.084     & 0.167    &  0.154                    \\ 
      CPU time                                   &   0.55     &  0.27    &  0.16                        \\ \hline
\end{tabular}
}
\caption{Samples with $32\%$ of outliers ($x$-outliers=10, $y$-outliers=6), $n=50$, $p=3$.
True: $\beta_{0}=0.0$, $\beta_{1}=1.20$, $\beta_{2}=-0.80$.} \label{table_7}
\end{table}

\section{Conclusions and Future Research}\label{sec_conclusions}
The PTS estimator can be considered as a procedure which instead of improving upon previous robust estimators such 
as LTS and MCD, it combines the information so obtained from them in a penalized manner, so as to simultaneously retain
their favorable robust  characteristics, while being efficient and identifying multiple masked outliers. Specifically, this
is achieved by using a robust scale and leverage from the LTS and MCD respectively.  Moreover, an efficient fast algorithm
which is based on a set of necessary conditions for local optimality is also presented. Extensive computational experiments
on a large set of instances, indicate that the proposed estimator ouperforms other robust estimators in the presence of 
all possible types of outliers. 

Future research  can be performed on both the estimator and the algorithm. With respect to the estimator, a different robust
scale $\hat{\sigma}$  could be investigated, and also it can be extended to treat data sets which contain a mixture of 
multilinear regression models. The algorithm could also be modified to handle massive data sets, which may have millions 
of observations, by utilizing the natural parallel structure of the algorithm. 

\bibliographystyle{plain}
\bibliography{robust}

\begin{thebibliography}{10}

\bibitem{AgoMar:98}
C.~Agostinelli and M.~Markatou.
\newblock A one-step robust estimator for regression based on the weighted
  likelihood reweighting scheme.
\newblock {\em Statistics and Probability Letters}, 37:342--350, 1998.

\bibitem{Agullo:01}
J.~{Agull\'{o}}.
\newblock New algorithms for computing the least trimmed squares regression
  estimator.
\newblock {\em Computational Statistics and Data Analysis}, 36:425--439, 2001.

\bibitem{Atkinson:94}
A.C. Atkinson.
\newblock Fast very robust methods for the detection of multiple outliers.
\newblock {\em Journal of the American Statistical Association}, 89:1329--1339,
  1994.

\bibitem{AtkChen:94}
A.C. Atkinson and T.-C. Cheng.
\newblock Computing least trimmed squares regression with the forward search.
\newblock {\em Statistics and Computing}, 9:251--263, 1999.

\bibitem{KaKuTu:93}
M.S. Bazaraa, H.D. Sherali, and C.M. Shetty.
\newblock {\em Nonlinear Programming: Theory and Algorithms}.
\newblock Wiley, 1993.

\bibitem{CoaHet:93}
C.W. Coakley and T.P. Hettmansperger.
\newblock A bounded influence, high breakdown, efficient regression estimator.
\newblock {\em Journal of the American Statistical Association}, 88:872--880,
  1993.

\bibitem{DonHub:83}
D.L. Donoho and P.J. Huber.
\newblock The notion of breakdown point.
\newblock In {\em A Festschrift for Erich Lehmann}, Belmont, CA, 1983.
  Wadsworth.

\bibitem{GenWilk:75}
J.F. Gentleman and M.B. Wilk.
\newblock Detecting outliers ii: Supplementing the direct analysis of
  residuals.
\newblock {\em Biometrics}, 31:387--410, 1975.

\bibitem{GerYoh:02}
D.~Gervini and V.J. Yohai.
\newblock A class of robust and fully efficient regression estimators.
\newblock {\em Annals of Statistics}, 30:583--616, 2002.

\bibitem{GilPad_a:02}
A.~Giloni and M.~Padberg.
\newblock Least trimmed squares regression, least median squares regression,
  and mathematical programming.
\newblock {\em Mathematical and Computer Modelling}, 35:1043--1060, 2002.

\bibitem{HadSim:93}
A.~S. Hadi and J.S. Simonoff.
\newblock Procedures for the identification of multiple outliers in linear
  models.
\newblock {\em Journal of the American Statistical Association}, 88:1264--1272,
  1993.

\bibitem{Hawkins:94}
D.M. Hawkins.
\newblock The feasible solution algorithm for least trimmed squares regression.
\newblock {\em Computational Statistics and Data Analysis}, 17:185--196, 1994.

\bibitem{HawBraKas:84}
D.M. Hawkins, D.~Bradu, and G.V. Kass.
\newblock Location of several outliers in multiple regression data using
  elemental sets.
\newblock {\em Technometrics}, 26:197--208, 1984.

\bibitem{HawOli:99}
D.M. Hawkins and D.J. Olive.
\newblock Improved feasible solution algorithms for high breakdown estimation.
\newblock {\em Computational Statistics and Data Analysis}, 30:1--11, 1999.

\bibitem{Hossjer:95}
O.~H{\"o}ssjer.
\newblock Exact computation of the least trimmed squares estimate in simple
  linear regression.
\newblock {\em Computational Statistics and Data Analysis}, 19:265--282, 1995.

\bibitem{Li:05}
L.M. Li.
\newblock An algorithm for computing exact least-trimmed squares estimate of
  simple linear regression with constraints.
\newblock {\em Computational Statistics and Data Analysis}, 48:717--734, 2005.

\bibitem{PenYoh:99}
D.~{Pe\~{n}a} and V.J. Yohai.
\newblock A fast procedure for outlier diagnostics in large regression
  problems.
\newblock {\em Journal of the American Statistical Association}, 94:434--445,
  1999.

\bibitem{PitRes02a}
L.S. Pitsoulis and M.G.C. Resende.
\newblock Greedy randomized adaptive search procedures.
\newblock In P.M. Pardalos and M.G.C. Resende, editors, {\em Handbook of
  Applied Optimization}, pages 168--183. Oxford University Press, 2002.

\bibitem{Rousseeuw:84}
P.J. Rousseeuw.
\newblock Least median of squares regression.
\newblock {\em Journal of the American Statistical Association}, 79:871--880,
  1984.

\bibitem{RouDrie:99}
P.J. Rousseeuw and K.~Van Driessen.
\newblock A fast algorithm for the minimum covariance determinant estimator.
\newblock {\em Technometrics}, 41:212--223, 1999.

\bibitem{RouDrie:06}
P.J. Rousseeuw and K.~Van Driessen.
\newblock Computing {LTS} regression for large data sets.
\newblock {\em Data Mining and Knowledge Discovery}, 12:29--45, 2006.

\bibitem{RouLer:87}
P.J. Rousseeuw and A.M. Leroy.
\newblock {\em Robust Regression and Outlier Detection}.
\newblock John Wiley, New York, 1987.

\bibitem{RouYoh:84}
P.J. Rousseeuw and V.J. Yohai.
\newblock Robust regression by means of s-estimators.
\newblock In {\em Robust and Nonlinear Time Series Analysis}, pages 256--272.
  Springer Verlag, 1984.

\bibitem{RouZom:90}
R.J. Rousseeuw and B.C.~Van Zomeren.
\newblock Unmasking multivariate outliers and leverage points.
\newblock {\em Journal of the American Statistical Association}, 85:633--639,
  1990.

\bibitem{BarYoh:06}
M.~Salibian-Barrera and V.J. Yohai.
\newblock A fast algorithm for s-regression estimates.
\newblock {\em Journal of Computational and Graphical Statistics},
  15(2):414--427, 2006.

\bibitem{Yohai:87}
V.J. Yohai.
\newblock High breakdown-point and high efficiency robust estimates for
  regression.
\newblock {\em Annals of Statistics}, 15:642--656, 1987.

\bibitem{YohZam:88}
V.J. Yohai and R.H. Zamar.
\newblock High breakdown point estimates of regression by means of the
  minimization of an efficient scale.
\newblock {\em Journal of the American Statistical Association}, 83:406--413,
  1988.

\bibitem{ZiouAv:05}
G.~Zioutas and A.~Avramidis.
\newblock Deleting outliers in robust regression with mixed integer
  programming.
\newblock {\em Acta Mathematicae Applicatae Sinica}, 21:323--334, 2005.

\bibitem{ZiAvPi:07}
G.~Zioutas, A.~Avramidis, and L.~Pitsoulis.
\newblock Penalized trimmed squares and a modification of support vectors for
  unmasking outliers in linear regression.
\newblock {\em REVSTAT}, 5:115--136, 2007.

\end{thebibliography}

\end{document}